\documentclass[aps,prl,twocolumn,amssymb,showpacs,groupedaddress]{revtex4-1}
\usepackage{graphicx}
\usepackage{amsmath}
\usepackage{changes}
\def\beq{\begin{equation}}
\def\eeq{\end{equation}}
\def\bea{\begin{eqnarray}}
\def\eea{\end{eqnarray}}

\begin{document}


\title{Persistence of perfect packing in twisted bundles of elastic filaments} 


\author{Andreea Panaitescu,$^\star$ Gregory M. Grason,$^\dag$ and Arshad Kudrolli$^\star$}
\affiliation{$^\star$Department of Physics, Clark University, Worcester, MA 01610,\\ 
$^\dag$Department of Polymer Science and Engineering, University of Massachusetts, Amherst, MA 01003}

\date{\today}

\begin{abstract}
It is generally understood that geometric frustration prevents maximal hexagonal packings in uniform filament bundles upon twist. We demonstrate that a hexagonal packed elastic filament bundle can preserve its order over a wide range of twist due to a subtle counteraction of geometric expansion with elastic contraction. Using x-ray scanning and by locating each filament in the bundle, we show the remarkable persistence of order even as the twist is increased well above $360$ degrees, by measuring the spatial correlation function across the bundle crosssection. We introduce a model which analyzes the combined effects of elasticity including filament stretching, and radial and hoop compression necessary to explain this generic preservation of order observed with Hookean filaments. 
\end{abstract}


\maketitle


Twisted filament bundle arrangements, common in soft matter and biological tissues, are a product of a subtle interplay of geometry, elasticity, and applied boundary conditions. It has been long recognized that a maximally-packed, hexagonally ordered bundle of filaments cannot be twisted without introducing disorder and finite gaps between neighbors~\cite{schwarz52}. Thus, arrangements of filaments to maximize packing or strength have been the subject of many studies motivated initially by the cable and textile industry and biology~\cite{freeston75,pan02,chouaieb06}, and more recently by the growing research on twisted self-assembly and functional materials~\cite{blair03,lima11,chopin13,hennecke13,haines16,frenzel17}. Significant insight has been gained in the case of bundles that achieve a uniform helical pitch~\cite{neukrich02,starostin06,olsen10}, in particular, the discovery that their 3D packing can be mapped to disk packing on a non-Euclidean surface with a Gaussian curvature which increases with twist~\cite{bruss12,grason15}. 

This geometric approach was recently tested with experiments using a twisted assembly of easily bendable but otherwise inextensible filaments which were observed to develop a constant helical rotation and outward expansion and rearrangement of packing with twist~\cite{panaitescu17}. Further, it was shown that the dual non-Euclidean surface can be approximated by a spherical cap with a curvature radius proportional to the helical pitch. Thus, defects are always introduced for sufficiently large twist because it is impossible to have a uniform spacing hexagonal lattice packing surface of non-zero Gaussian curvature. While the effect of geometric frustration to disrupt 
order in self-assembled states is well-studied~\cite{bowick09,gibaud12,meng14,grason16,lenz17}, the role of the elasticity of the constituent elements themselves on the response of an organized structure to frustration has received little attention.

Here, we show that notwithstanding the tendencies of geometric frustration to disrupt packing in uniform diameter filament bundles, ``perfect" hexagonal packing is in fact well maintained in twisted bundles of {\it elastic} filaments. This persistence of order arises from a subtle combination of the decrease in crosssectional area of stretched elastic filaments along with their ability to mutually compress laterally, which nearly perfectly balances the expansile tendencies of ``positive-curvature" geometry imposed by twist. Unlike proposed ordered lattice packings with parallel perfectly bendable, yet incompressible, filaments~\cite{starostin06}, we find that any isotropic filament composed of Hookean elastomers will preserve hexagonal order to a surprising degree over a wide range of twist that results in non-parallel filaments.

\begin{figure}
\begin{center}
\includegraphics[width=0.45\textwidth]{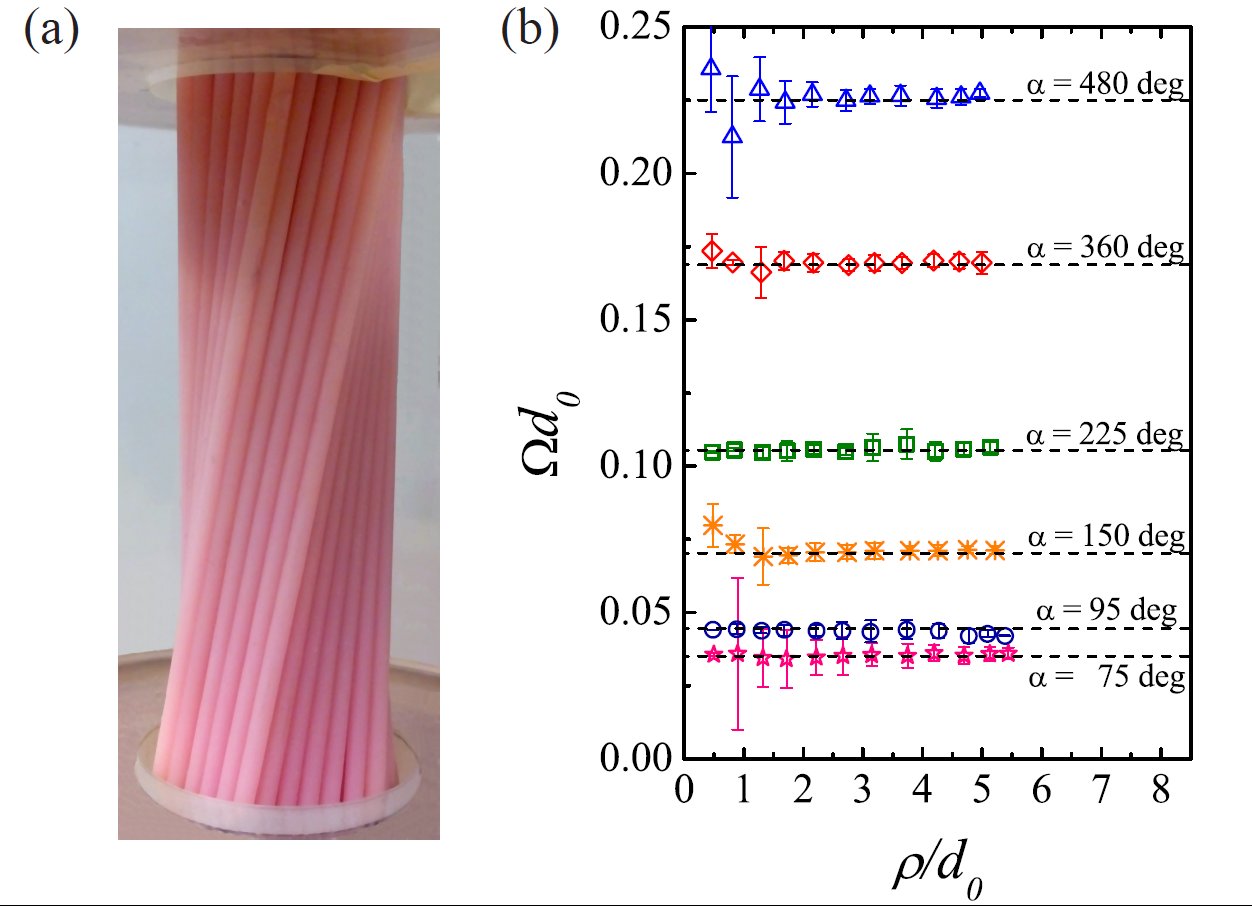}
\end{center}
\caption{(a) Image of a twisted elastic filament bundle. (b) The measured helical angle $\Omega$ as a function of distance $\rho$ from the axis of rotation for each filament is roughly constant. The horizontal dashed lines correspond to $\Omega = \alpha/ H$.}
\label{fig:image}
\end{figure} 

Filaments composed of vinylpolysiloxane were first manufactured using cylindrical molds with diameter $d_0 = 2.6 \pm 0.05$\,mm, Young's modulus $E = 220 \pm 10$\,MPa, and Poisson's ratio $\nu = 0.4 \pm 0.02$~\cite{supdoc}. A hexagonal bundle with 102 filaments is then assembled, with each filament clamped to a hexagonal template at its two ends separated by a distance $H = 96$\,mm. The filaments are then twisted by rotating the top clamp through an angle $\alpha$ resulting in the helical metastructure shown in Fig.~\ref{fig:image}(a). We then use a micro x-ray Computed Tomography instrument to scan the filament bundle and identify their 3D positions using the image processing toolbox in MATLAB. Then, we determine the radial distance from the axis of rotation $\rho$ and helical pitch $2 \pi/\Omega$ of each filament by fitting a helix to the measured data points as in our previous work~\cite{panaitescu17}. The measured $\Omega$ as a function of $\rho$ is plotted in Fig.~\ref{fig:image}(b) shows that inverse pitch remains constant to within small fluctuations for each $\alpha$. Further, the measured $\Omega$ is found to increases with twist, consistent with $\Omega = \alpha/H$.

\begin{figure}
\begin{center}
\includegraphics[width=0.45\textwidth]{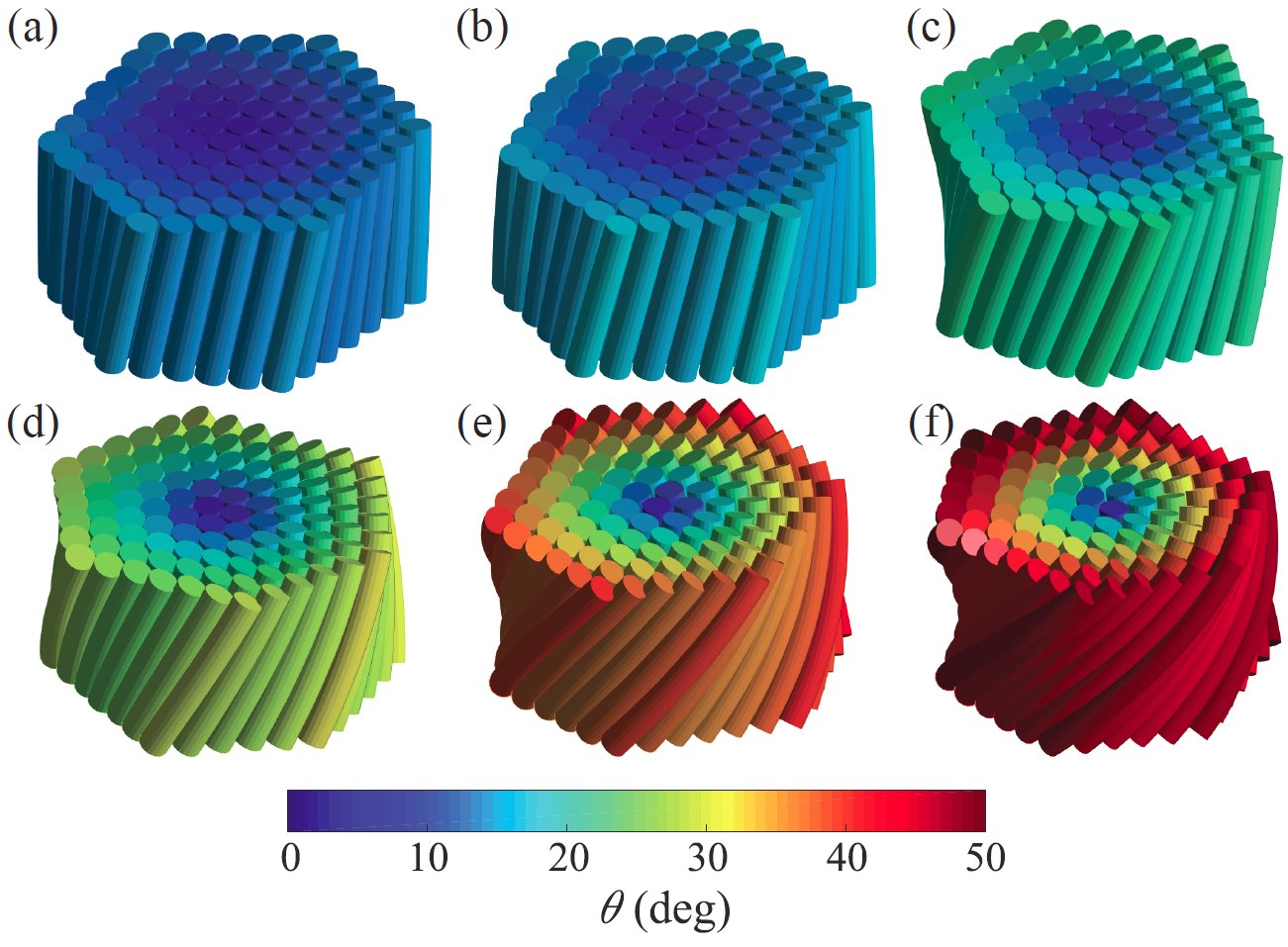}
\end{center}
\caption{(a-f) The reconstructed bundle shapes obtained with micro x-ray scanning with a color corresponding to the filament inclination angle $\theta$. (a) $\alpha = 75^0$, (b) $\alpha = 95^0$, (c) $\alpha = 150^0$, (d) $\alpha = 225^0$, (e) $\alpha = 360^0$, and (f) $\alpha = 480^0$.  }
\label{fig:3d}
\end{figure} 

\begin{figure}[t]
\begin{center}
\includegraphics[width=0.45\textwidth]{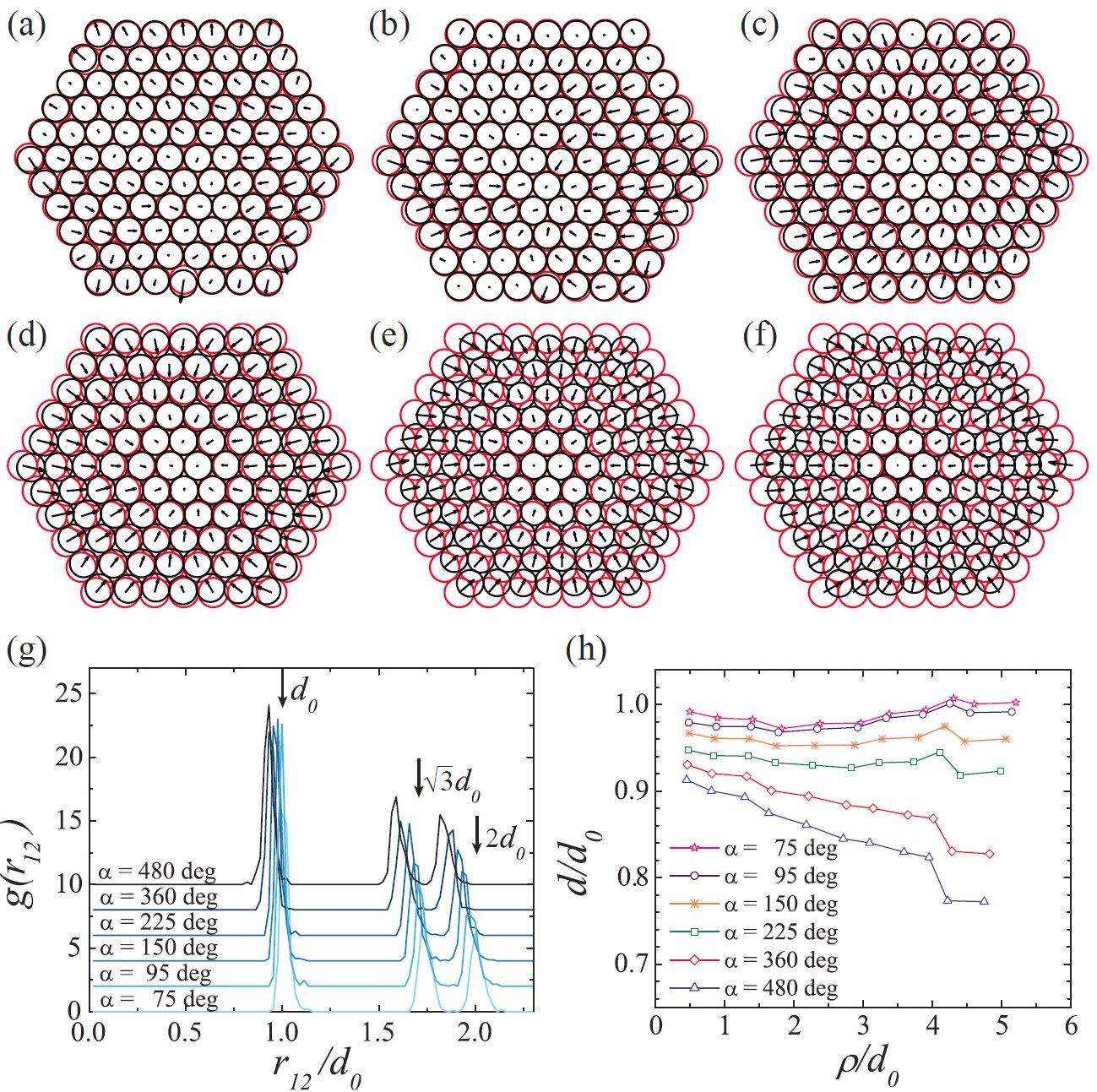}
\end{center}
\caption{(a-f) Filament displacement in a horizontal cross section as a function of increasing twist for $\alpha$ shown in Fig.~\ref{fig:3d}(a-f) using initial close contact hexagonal lattice as a reference. The arrows (displacements) are normalized within each sub-figure. (g) The two-point spatial correlation function $g(r)$, measured in the planar transect, quantitatively shows the persistence of hexagonal order. (h) The measured $d$ is observed to vary systematically as a function of $\rho$ and $\alpha$ (see text).  
}
\label{fig:cross}
\end{figure}

Renderings of the packings for increasing twist are shown in Fig.~\ref{fig:3d}(a-f) in a central section of the bundle away from the direct influence of the end clamps. The inclination angles of the filaments with respect to the rotation axis, given by $\theta = \cos^{-1}{(1/\sqrt{1 + (\Omega \rho)^2})}$, are shown using the given color map. One observes that the bundle crosssections appear remarkably ordered with increasing twist as $\theta$ increases from $0^o$ at the core to above $50^o$ at the outer boundary of the bundle even for the highest twist. To examine this structure more clearly, we plot the position of each filament in a transect (red/gray) superimposed on the corresponding hexagonal lattice packing of the filaments (black) in Fig.~\ref{fig:cross}(a-f). Further, arrows corresponding to the direction that the filaments move is also drawn to visualize the rearrangements that occur within each transact as the bundle is twisted. We observe that the relative symmetry of the bundle is remarkably well preserved over a wide range of twist.  

To quantify the inter-filament order, we plot the two-point spatial correlation function $g(r_{12})$ in Fig.~\ref{fig:cross}(g) as a function of distance $r_{12}$ between filament centers in the transect normalized by $d_0$. We clearly observe all the three distinct peaks corresponding to a hexagonal packing over the interval plotted, with the peaks shifting slightly inward with increasing $\alpha$. To understand this deformation at the filament level, we measure and plot the mean diameter $d$ of the filaments as a function of $\rho$ for the various twisted bundles in Fig.~\ref{fig:cross}(h). Here, $d$ is obtained by measuring the average center to center perpendicular distance between all neighbor filaments to account for the fact that the filament sections become increasingly polygonal as  filaments are increasing squeezed together. We observe an overall decrease in $d$ with $\alpha$ and $\rho$.  Thus, although the filament diameters themselves change individually, their collective  packing in the planar transect remains ordered as quantified by the unchanged peak heights and widths in Fig.~\ref{fig:cross}(g). 

\begin{figure}
\begin{center}
\includegraphics[width=0.45\textwidth]{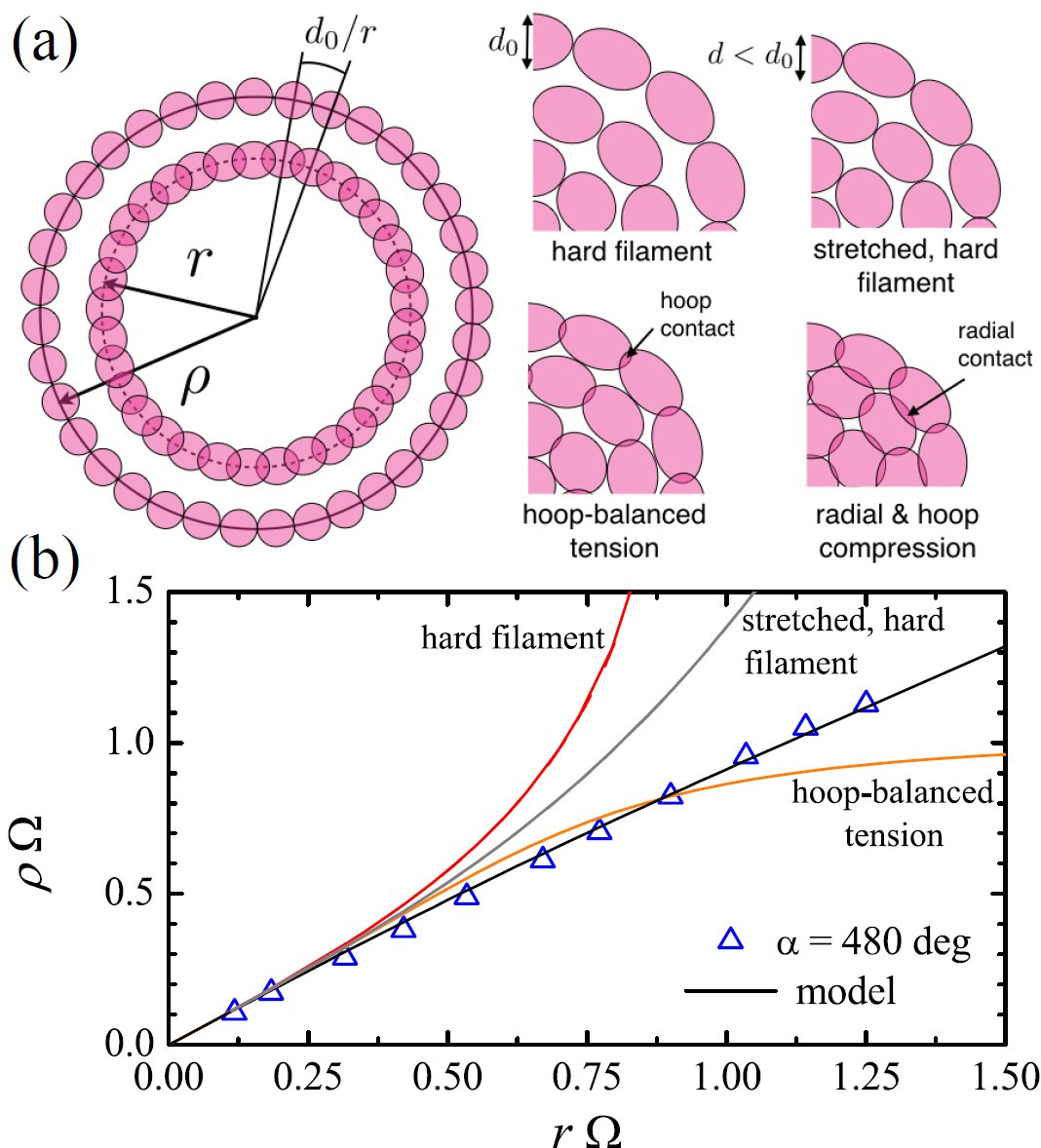}
\end{center}
\caption{(a) A schematic representation of filaments before and after twist is applied along with graphical representation of the various geometric and elastic interactions between the filaments.  (b) The model is observed to capture the measured $\rho$. The contribution of the various counteracting components of the model are also plotted.  
}
\label{fig:rhor}
\end{figure} 

The preservation of hexagonal order in the transect shown in Fig.~\ref{fig:cross} is consistent with an approximately uniform radial contract, $\rho \propto r$, which is quite unlike expansile and non-linear deformation profile observed in twisted bundles of inextensisible, or ``hard", filaments reported previously~\cite{freeston75,bruss12,wang15,panaitescu17}.  To understand the observed structure, we now construct a model that integrates four mechanical and geometrical effects of imposed twist:  i) the geometric frustration of inter-filament spacing; ii) the diameter contraction of stretched filaments; iii) the inward force on tensed, helical filaments; and iv) the outward pressure sustained by inter-filament contacts with both {\it azimuthal} and {\it radial} neighbors in the bundle.  To better understand these individual effects which are illustrated in Fig.~\ref{fig:rhor}(a) on the collective response, we describe each of their ingredients, in turn, and consider there cumulative effect on the ultimate model that integrates all of them.  For simplicity, we focus on the axisymmetric deformation of filament positions and seek the mapping between the initial radial position $r$ of filaments in a parallel, untwisted bundle and their final position $\rho$ upon imposing twist $\Omega$.

We begin by describing the generation of geometric frustration by twist.  Helical tilt relative to the central axis of the bundle increases with radius as,
\begin{equation}
\theta(\rho) = \arctan (\Omega \rho) ,
\end{equation}
which alters the relationship between center-to-center spacing in the planar transect ${\bf \Delta}$ and the true spacing ${\bf \Delta}_{\perp}={\bf \Delta}- {\bf t} \big({\bf t} \cdot {\bf \Delta})$, where ${\bf t} = \cos \theta~ \hat{z} + \sin \theta ~\hat{\phi}$, is the local tangent at $\rho$, and $z$ and $\phi$ are respective vertical and azimuthal coordinates.  Twist effectively subjects filaments at a given radius to geometric compression with their neighbors in the same hoop relative to their parallel state~\cite{grason15}, while spacing of radial neighbors is unperturbed by this tilt pattern.  If the filaments were to maintain constant crossectional diameters, i.e. the {\it hard filament} (HF) assumption, then the final spacing between filaments in the compressed hoop remains constant. Thus, $ 2 \pi \rho \cos \theta / N (\rho) = d_0$, where $N(\rho) = 2 \pi r/d_0$ is the number of filaments in the hoop before and after twist.  The HF constraint of constant hoop spacing then requires the radial deformation,
\begin{equation}
\label{eq: hard}
\rho_{HF}(r) = \frac{ r}{\sqrt{1-(\Omega r)^2 }} ,
\end{equation}
which is plotted in Fig.~\ref{fig:rhor}(b) and shows that each hoop of filaments expands progressively in order to avoid azimuthal overlap imposed by twist~\cite{bohr11}. This dilatory response to twist was indeed observed for the case of unstretchable and relatively incompressible rods previously~\cite{panaitescu17}, but clearly fails to capture the collective response for the present case of stretchable, elastic filaments plotted for $\alpha = 480^o$ which increases almost linearly consistent with the relatively small contraction overall observed even over this large twist.  

We now consider the effect of {\it stretching} of elastic filaments to reduce their crosssectional diameter. Clamping of filaments at their ends stretches their lengths to $L = H/\cos \theta$, and assuming perfect incompressibility for simplicity, we have 
 $\pi d_0^2 H/4 = \pi d^2 L/4$, and thus the contracted diameter 
$d(\rho) = d_0 \cos^{1/2} \theta\,$.
Note that the effect of diameter contraction increases with radial position in bundle, and as such, it tends to counteract, somewhat, the effect of geometrically-imposed compression along the hoops discussed above.  Taking into account this reduced diameter at $\rho$, but retaining the assumption that inter-filament contact does not alter their narrowed circular crosssections, which one may call a {\it stretched, hard filament}  (SHF) approximation, leads to a modified deformation profile,
\begin{equation}
\label{eq: SHF}
\rho_{\rm SHF} (r)= r \sqrt{\sqrt{1+(\Omega r)^4/4}+(\Omega r)^2/2} \,,
\end{equation}
which is compared against the data and eq. (\ref{eq: hard}) in Fig.~\ref{fig:rhor}(b).  Indeed the elastic diameter contraction of outer filaments leads to a reduction in the expansion required to avoid neighbor overlaps relative to the assumption of constant diameter (i.e., $\rho_{\rm SHF} < \rho_{\rm HF}$). However, the profile remain expansile, or $\rho_{\rm SHF}  > r$, in direct contrast with observed deformations.  

We now incorporate the effect of tension in stretched filaments to drive radial compression in twisted bundles, a drive that is balanced by the elastic cost of inter-filament compression.  For clarity of illustration, we first approximate this mechanical equilibrium by considering only the elastic compression of hoops at a given radius described above, neglecting momentarily, radial compression of distinct hoops. Filament stretching leads to tension $T = E \pi d_0^2 (\sec \theta -1)$, where $E$ is the modulus, and due to the curved geometry of filaments, sections of the filaments experience an inward (radial) force per unit length $F_{tense}/L = - T \kappa$ where $\kappa = \sin^2 \theta/\rho$ is the local curvature.  This inward force must be balanced by the elastic compression of lateral neighbors in the bundle.   Azimuthal neighbors interpenetrate by an amount $\delta_\phi = d(\rho)-\Delta_\phi \cos \theta $, leading to a (Herztian) contact force (per unit length) between neighbors $F_{Hertz}/L= E \pi \delta_\phi/3$~\cite{johnson87}, which due to the radial geometry of hoops gives rise to an outward force $F_{hoop}= 2 F_{Hertz} \Delta_\phi/\rho$.  Balancing these elastic forces ($F_{tense} = F_{hoop}$) yields the equilibrium condition for this {\it hoop-balanced tension} (HBT) approximation,
\begin{equation}
\label{eq: hoopbalance}
 \sin^2 \theta (\sec \theta - 1) = \frac{2 r}{3 \rho } \Big(  \frac{r  }{\rho} \cos \theta - \cos ^{1/2} \theta \Big) .
\end{equation}
The solution $\rho_{\rm HBT}(r)$ by solving eq. (\ref{eq: hoopbalance}) is also plotted in Fig.~\ref{fig:rhor}(b), showing that effect of tension is indeed to drive inward compression relative to the combined effect geometric frustration and diameter contraction alone, described by eq. (\ref{eq: SHF}).  While near the core, the prediction is slightly expansile, $\rho_{\rm HBT} < \rho_{\rm SHF}$, indicating that tension elastic compression of neighbor diameters is smaller there.  For larger radii, $\Omega r \gtrsim 0.70$, the HBT crosses over into a contractile regime where $\rho_{HBT}<r$.  While the contractile nature of the deformation is qualitatively consistent with experiments, the extent of contraction predicted by the ``hoop only" mechanical balance grossly overestimates this contraction at large radii~\cite{supdoc}. 

Thus, we finally discuss the model that incorporates all of the above effects as well as the compression of filaments in both the {\it azimuthal} and {\it radial} distances. It is most convenient to analyze this model in the continuum limit (where $d_0 \ll R$) and in terms of the energy-minimizing state of the total elastic energy, ${\cal E}$ of a twisted bundle of total initial radius $R$,
\begin{multline}
\label{eq: Aspen}
{\cal E}  =  H E \Phi_0 \int_0^R dr~r \bigg\{\big(\sec \theta -1 \big)^2 \\ + \frac{3}{2} \Big[\big(\frac{\rho}{r} \cos \theta -\cos^{1/2} \theta \big)^2 + \big(\rho' - \cos^{1/2}\theta \big)^2 \Big] \bigg\} ,
\end{multline}
where $\Phi_0$ and $R$ are the respective area fraction and radius of the initially untwisted array.  As described in detail in the Supporting Materials~\cite{supdoc}, the first term in the integrand describes elastic cost of filament stretching by imposed twist, while the second and third terms represent the elastic energy stored in Hertzian compression of azimuthal and radial neighbors, respectively.  The equilibrium radial profile, $\rho_{eq}(r)$, follows from solving the Euler-Lagrange equation for $\frac{ \delta {\cal E} }{ \delta \rho(r) }=0$, subject to the condition of vanishing radial compression, $\rho'= \cos^{1/2} \theta$, at the bundle surface, $r=R$. For $\Omega R \leq 1.19$, the equilibrium profile includes an inner boundary at $r_*=R$ where radial compression vanishes, and  $\rho_{eq}(r)$ matches (continuously) onto the  $\rho_{\rm HBC}(r)$ for $r\leq r_*$ (see Supporting Materials for detailed profiles~\cite{supdoc}.)  

The equilibrium profile, $\rho_{eq}(r)$, is shown in Fig.~\ref{fig:rhor}(b), for a case of $\alpha = 480^\circ$, showing good quantitative and qualitative agreement with the experimental deformation.  In particular we find, relative to the HBT approximation, that additional elastic cost of radial compression included in eq. (\ref{eq: Aspen}) has two critical effects.  First, it significantly {\it reduces} the degree of radial compression generated to tensile contraction at the outside of the bundle.  Second, radial contact between successive hoops facilitates transfer of compression from outer hoops into the central region of the bundle, a feature missing from the ``hoops only" approximation.  These combined effects give rise to the nearly linear, compressional displacement pattern (e.g. $\rho \propto r$), which simply rescales the neighbor spacing, without disrupting the hexagonal order of the transect.

\begin{figure}
\begin{center}
\includegraphics[width=0.48\textwidth]{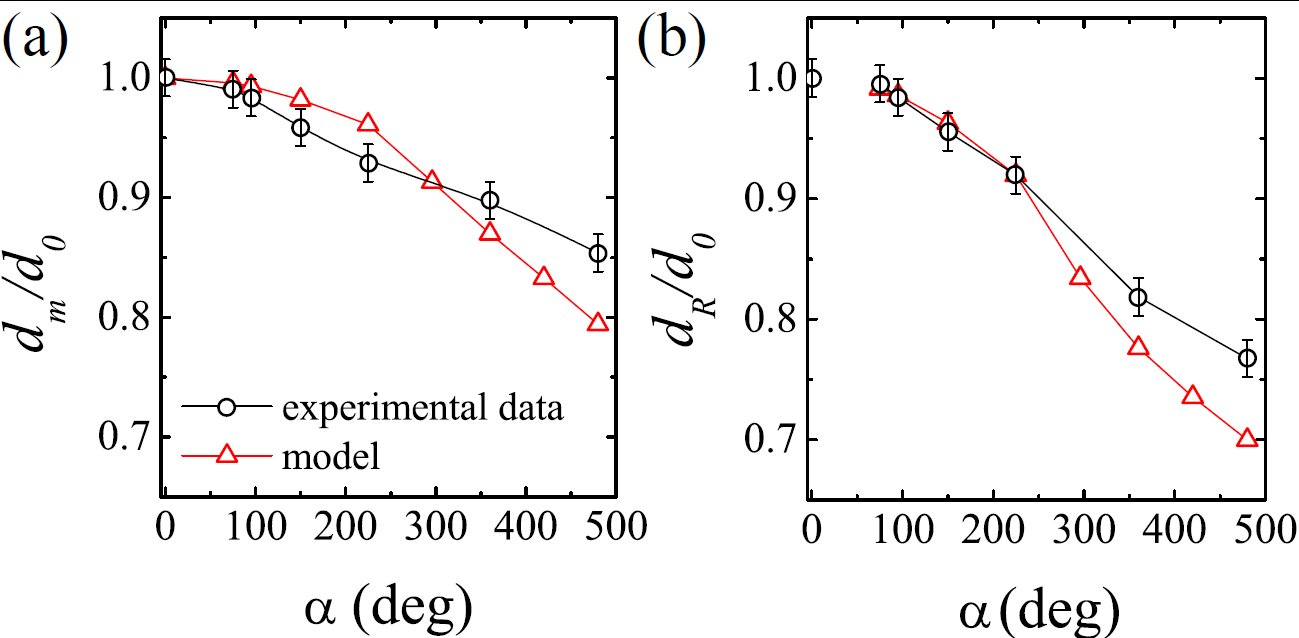}
\end{center}
\caption{Comparison of the measured and estimated mean filament diameters in the bundle $d_m$ (a) and the mean diameters of the outermost filaments $d_R$ (b) as a function of twist.}
\label{fig:dia}
\end{figure}

We plot the mean diameter of filaments $d_m$ in Fig.~\ref{fig:dia}(a) and the diameter of outermost six filaments $d_R$ in Fig.~\ref{fig:dia}(b) along with the computed curves corresponding to the model. For both quantities, the local diameter is computed by averaging over compression in the radial and hoop directions, i.e. $d=d_0\cos^{1/2}\theta - (\delta_r+\delta_\phi)/2$, where $\delta_r$ and $\delta_\phi$ are the elastic indentation of radial and azimuthal neighbors, respectively.  Notwithstanding the simplifying assumptions of perfect filament incompressibility axisymmetry of the packing, we find overall agreement with the measured filament diameters over the range of twists studied.  

In conclusion, we have demonstrated that a close packed elastic filament bundle can retain hexagonal order under twist in contrast with naive expectations that twist must obstruct hexagonal order due to geometric frustration. A simple contraction of the filaments due to stretching alone is not sufficient to offset the inherent geometric frustration imposed by the expansion of the filament bundle with twist. Rather, compression of the filaments due to the further development of mechanical stresses in the radial and azimuthal direction are necessary to explain the relative invariance of the observed bundle structure with twist. Remarkably, the observed ordered structures constructed out of linear elastic materials are relatively insensitive to the elastic modulus according to our calculations, showing the robustness of the phenomena, and their wide applicability. Finally, our work illustrates how elasticity can alleviate the frustration imposed by geometrical constraints due to collective deformation of multi-element architectures.

\begin{acknowledgments}
This work was supported by the National Science Foundation under grant numbers DMR 1508186 (A.P. and A.K.) and DMR 1608862 (G.G.), and was performed in part at the Aspen Center for Physics, which is supported by National Science Foundation grant PHY-1607611.
\end{acknowledgments}

\end{document}